\documentstyle[twoside,fleqn,espcrc2]{article}

\newskip\humongous \humongous=0pt plus 1000pt minus 1000pt
\def\caja{\mathsurround=0pt}
\def\eqalign#1{\,\vcenter{\openup1\jot \caja
        \ialign{\strut \hfil$\displaystyle{##}$&$
        \displaystyle{{}##}$\hfil\crcr#1\crcr}}\,}
\newif\ifdtup

\def\eqright #1\cr{\noalign{\hfill$\displaystyle{{}#1}$}}
\def\eqleft #1\cr{\noalign{\noindent$\displaystyle{{}#1}$\hfill}}

\def\oldreffmt#1{\rlap{[#1]} \hbox to 2\parindent{}}

\def\figfmt#1{\rlap{Figure {#1}} \hbox to 1in{}}

\def\sectioneq{\def\theequation{\thesection.\arabic{equation}}{\let
\holdsection=\section\def\section{\setcounter{equation}{0}\holdsection}}}%
\def\sectiontab{\def\thetable{\thesection.\arabic{table}}{\let
\holdsection=\section\def\section{\setcounter{table}{0}\holdsection}}}%
\def\sectionfig{\def\thefigure{\thesection.\arabic{figure}}{\let
\holdsection=\section\def\section{\setcounter{figure}{0}\holdsection}}}%

\newcounter{holdequation}

\def\begineq #1\endeq{$$ \refstepcounter{equation}\eqalign{#1}\eqno
	(\theequation) $$}
\def\contlimit{\,{\hbox{$\longrightarrow$}\kern-1.8em\lower1ex
\hbox{${\scriptstyle (a\rightarrow0)}$}}\,}
\def\centeron#1#2{{\setbox0=\hbox{#1}\setbox1=\hbox{#2}\ifdim
\wd1>\wd0\kern.5\wd1\kern-.5\wd0\fi
\copy0\kern-.5\wd0\kern-.5\wd1\copy1\ifdim\wd0>\wd1
\kern.5\wd0\kern-.5\wd1\fi}}
\def\centerover#1#2{\centeron{#1}{\setbox0=\hbox{#1}\setbox
1=\hbox{#2}\raise\ht0\hbox{\raise\dp1\hbox{\copy1}}}}
\def\centerunder#1#2{\centeron{#1}{\setbox0=\hbox{#1}\setbox
1=\hbox{#2}\lower\dp0\hbox{\lower\ht1\hbox{\copy1}}}}
\def\lsim{\;\centeron{\raise.35ex\hbox{$<$}}{\lower.65ex\hbox
{$\sim$}}\;}
\def\gsim{\;\centeron{\raise.35ex\hbox{$>$}}{\lower.65ex\hbox
{$\sim$}}\;}
\def\st#1{\centeron{$#1$}{$/$}}

\def\super#1{\ifmmode \hbox{\textsuper{#1}}\else\textsuper{#1}\fi}
\def\textsuper#1{\newcount\holdspacefactor\holdspacefactor=\spacefactor
$^{#1}$\spacefactor=\holdspacefactor}

\def\getcite#1,{\advance\citenumber by1
\ifnum\citenumber=1
\ref{#1}\let\next=\getcite\else\ifx#1@\let\next=\relax
\else ,\ref{#1}\let\next=\getcite\fi\fi\next}
\def\nskip#1{\vglue-\baselineskip\vglue#1\vglue-\parskip\noindent}

\def\upon #1/#2 {{\textstyle{#1\over #2}}}
\relax

\font\ssi=cmssi12                           
\def\emph{\ssi}

\newcommand{\AmS}{{\protect\the\textfont2
  A\kern-.1667em\lower.5ex\hbox{M}\kern-.125emS}}

\title{\nskip{-2.95truecm} \rightline{
\vbox{\normalsize
\halign{&## \hfil\cr
&ANL-HEP-CP-92-103\cr
&November 10, 1992\cr}
}}\nskip {2.70truecm}%
A streamlined method for chiral fermions on the lattice\thanks{Talk
presented by G.~Bodwin at the Lattice~92 Symposium, Amsterdam, The
Netherlands, Sept. 15--19, 1992.}}

\author{Geoffrey T. Bodwin\address{Argonne National Laboratory, 9700 S. Cass
Ave.,\\ Argonne, IL 60439,USA}%
       \thanks{Work supported by the U.S. Department
of Energy, Division of High Energy Physics, Contract W-31-109-ENG-38.}
        and
        Eve V. Kov\'acs\address{Fermi National Accelerator Laboratory, P. O.
Box 500,\\Batavia, IL 60510, USA}}

\begin{document}

\begin{abstract}
We discuss the use of renormalization counterterms to restore the chiral
gauge symmetry in a lattice theory of Wilson fermions. We show that a
large class of counterterms can be implemented automatically by making a
simple modification to the fermion determinant.
\end{abstract}

\maketitle

Some time ago we presented a lattice method for chiral gauge theories
that involves the introduction of auxiliary Dirac species
\cite{lat90,pf91}.  Here we elaborate on an alternative approach
\cite{pf91} that achieves the effects of the auxiliary species through a
direct modification of the fermion determinant.  This alternative method
has the advantage that the computational algorithm is simpler, involving
two determinants instead of three. It also eliminates ambiguous square
roots of determinants that arise in the previous method.

\section{BASIC STRATEGY}

Our approach is similar in general philosophy to that of the Rome group
\cite{borrelli-et-al}. However, as we shall see, it differs
significantly in detail.

We begin by introducing a {\it Dirac} particle via the ``naive''
lattice action:
\begin{eqnarray}
\lefteqn{S_N=a^4\sum_{x,\mu}\overline\psi(x)\gamma_\mu{1\over
2a}[\psi(x+a_\mu)-\psi(x-a_\mu)].}\nonumber\\
&&
\end{eqnarray}
However, we couple the gauge field only to the part of the Dirac field
which, in the continuum limit of the action, would be the left-handed
component:
\begin{eqnarray}
\lefteqn{S_{NI}=a^4\sum_{x,\mu}\overline\psi(x)\gamma_\mu P_L
{1\over 2a}
[(U_\mu(x)-1)\psi(x+a_\mu)}\nonumber\\
&&\hskip 1.0cm -(U_\mu^\dagger(x-a_\mu)-1)\psi(x-a_\mu)],
\end{eqnarray}
where $P_{R/L}=(1/2)(1\pm \gamma_5)$.
The Feynman propagator corresponding to the naive action is
\begin{equation}
S_F^N(p_\mu)=[(1/a) \sum_\mu \gamma_\mu \sin(p_\mu a)]^{-1},
\end{equation}
which, in addition to the usual pole at $p=0$, has extra poles when one
or more momentum components are equal to $\pi/a$.  It can be seen
that half of the poles have positive chiral charge and half have
negative chiral charge \cite{doubling}, so, contrary to our initial
expectation, this doubling phenomenon leads to gauge-field couplings to
both left- and right-handed species.

We follow the standard approach of eliminating the doublers by including
a Wilson mass term \cite{wilson} in the action:
\begin{eqnarray}
\lefteqn{S_W=a^4\sum_{x,\mu}
\overline\psi(x)
{1\over 2a}[\psi(x+a_\mu)}\nonumber\\
&&\hskip 2.0cm+\psi(x-a_\mu)-2\psi(x)].
\end{eqnarray}
We can gauge the Wilson term by adding to the action
\begin{eqnarray}
\lefteqn{S_{WI}=a^4\sum_{x,\mu}\overline\psi(x) {1\over 2a}
[(U_\mu(x)-1)\psi(x+a_\mu)}\nonumber\\
&&\hskip1.0cm +(U_\mu^\dagger(x-a_\mu)-1)\psi(x-a_\mu)
].\label{gauge-wilson}
\end{eqnarray}
(As we shall see, it may sometimes be convenient to drop this coupling
of the Wilson term to the gauge field.) Now the propagator has a pole
only at $p=0$:
\begin{eqnarray}
\lefteqn{S_F^W=\{(1/a)\sum_\mu\gamma_\mu\sin(p_\mu a)}\nonumber\\
&&\hskip 1.0cm
+(2/a)\sum_\mu[1-\cos(p_\mu a)]\}^{-1},\label{prop-wilson}
\end{eqnarray}
which would seem to leave us, as desired, with a single Dirac particle
with only left-handed couplings to the gauge field.  Unfortunately, the
Wilson terms $S_W$ and $S_{WI}$, having the Dirac structures of masses,
break the chiral gauge invariance and couple the right-handed component
of the Dirac field back into the theory. Specifically, the difficulty is
that $\gamma_5$ commutes, rather than anticommutes, with the (identity)
Dirac matrices in $S_W$ and $S_{WI}$.  As a consequence, the chiral
gauge current is no longer conserved.

Such violations of chiral current conservation are unacceptable in a
chiral gauge theory since they jeopardize the decoupling of ghost fields
and, hence, unitarity.  Furthermore, current conservation is an
important ingredient in the standard renormalization program.  Without
it, there is an explosion of new counterterms, whose coefficients must
each be tuned in order to obtain a satisfactory theory. For example, in
the absence of current conservation, the vacuum polarization can
generate a quadratically divergent gauge-boson mass, the light-by-light
graph requires counterterms, Lorentz-noncovariant counterterms can
arise on the lattice, and, in non-Abelian theories, the
fermion--gauge-boson coupling can become different from the triple
gauge-boson coupling.

A key idea in our proposal (and in that of the Rome group), is that, by
tuning a suitable set of counterterms, one can restore chiral current
conservation in the continuum limit.  A heuristic argument in support of
this idea is the following.  We can regard the lattice formulation as a
UV regularization of the theory.  By definition, the difference between
the lattice regularization and any other UV regularization resides at
large loop momentum ($p_\mu\sim 1/a$). Because the Wilson term
eliminates the poles at $p_\mu=\pi/a$, large Euclidean loop momentum
implies that propagators are far off their mass shells. Then the
corresponding subdiagram is equivalent to a local interaction. Thus, if
there exists a satisfactory UV regularization of the chiral theory (that
is, one that respects the chiral gauge symmetry), then it is equivalent
to the Wilson lattice regularization plus local counterterms.

Therefore, we attempt to restore the chiral current conservation for the
Wilson action by adding to it local counterterms. If this procedure
succeeds, then the resulting theory is unique, up to a coupling-constant
renormalization.  The reason is that one has freedom only to alter the
coefficients of the gauge-invariant counterterms, and those counterterms
correspond to coupling-constant renormalization.

\section{THE STREAMLINED METHOD}

\subsection{General Considerations}

The counterterms of concern to us correspond to local parts of the
divergent subgraphs involving the fermion.  In four dimensions, these
subgraphs are the closed fermion loops with up to four external gauge
bosons, the fermion self-energy correction, and the fermion--gauge-boson
vertex correction. In general, there are counterterms corresponding to
every operator that is consistent with the symmetries of the lattice
theory and has dimension less than or equal to four.  For the gauged
Wilson theory, there are eleven such counterterms in the Abelian case
and more in the non-Abelian case. Four of these can be absorbed into two
additional coupling constant renormalizations, but the rest must be
dealt with separately.

Clearly it would be awkward to tune so many coefficients in simulation.
Therefore, we will try to find rules for computation that {\it
automatically} implement at least some of the counterterms. Our approach
will be to modify amplitudes in ways that correspond to adding local
contributions, with the goal of producing a final expression that respects
conservation of the chiral gauge current.

\subsection{Closed Loops}

Let us focus first on the divergent subgraphs involving closed fermion
loops.  We set aside for now the self-energy and vertex corrections.

By examining the Feynman identity, we can see at the graphical level how
the violations of current conservation occur.  For simplicity, we give
only a schematic form, which exhibits the essential features of the full
lattice expression:
\begin{equation}
\st k P_L =(\st p+\st k +M)
P_L - P_R(\st p+M)+M\gamma_5.
\label{feynman}
\end{equation}
On the right side of (\ref{feynman}), the first term cancels a
propagator on the left and the second term cancels a propagator on the
right.  If we were to apply the Feynman identity to a set of diagrams
containing all permutations of the fermion--gauge-field vertices, then
the contributions of the first two terms on the right side of
(\ref{feynman}) would exhibit the usual pair-wise cancellations that
appear in the textbook proofs of current conservation.  (In the case of
the lattice theory, a few complications, which are irrelevant for our
purposes, arise because of seagull vertices.)  However, the
contributions corresponding to the last term of (\ref{feynman}) would
remain and would correspond to a violation of chiral current
conservation.

The last term in (\ref{feynman}) appears because $\gamma_5$ commutes
with $M$.  This suggests that we try to restore lattice current
conservation by modifying the computational rules in such a way that
$\gamma_5$ effectively anticommutes with $M_W$.  Such a modification
would, of course, change the amplitude.  However, the change would be
proportional to $M_W$.  Now $M_W=\sum_\mu (2/a)[1-\cos(p_\mu a)]$
vanishes as $a\rightarrow 0$ unless $p_\mu\sim 1/a$.  Thus, such a
change in the amplitude is a purely local contribution in the continuum
limit and corresponds to the addition of a local counterterm to the
action. Since loop momenta of order $1/a$ can give important
contributions only in divergent subdiagrams, this modification would
leave the continuum limits of convergent subdiagrams unchanged. A
similar procedure is often used in the continuum in dealing with
dimensionally regulated graphs involving $\gamma_5$.  There the
prescription is to anticommute $\gamma_5$'s {\it before} continuing away
from  $d=4$.

We can exploit this anticommutation trick in order to eliminate the
$\gamma_5$'s in closed fermion loops in all terms containing an even
number of $\gamma_5$'s.  We anti-commute the $\gamma_5$'s through all
gamma matrices {\it and} through the Wilson mass.  Schematically, we
have
\begin{eqnarray}
\lefteqn{\left[\ldots \gamma_\mu P_L
{1\over \st p_1+M} \gamma_\nu P_L
{1\over \st p_2+M}\gamma_\rho P_L
\right]_{\vtop{\openup-1\jot
\halign{#\hfil\cr even no.\cr of $\gamma_5$'s\cr}}}}\nonumber\\
&&\longrightarrow \ldots\gamma_\mu{1\over \st p_1+M}\gamma_\nu{1\over \st
p_2+M}\gamma_\rho \left({1\over 2}\right).\label{even}
\end{eqnarray}
(Vertices arising from (\ref{gauge-wilson}) complicate the analysis
slightly, but do not change the conclusions.) The resulting expression
is vectorlike, so the gauge field couples to a conserved current, as
required.  Because of the factor $1/2$ on the right side of
(\ref{even}), such contributions correspond to the {\it square root} of
the determinant of the Wilson-Dirac operator for a fermion with
vector-like couplings to the gauge field. (The action is given by
$S_{N}+S_{NI}+S_{W}+S_{WI}$ with $P_L\rightarrow 1$.)

We can implement this trick conveniently in simulations by noting
that
\begin{eqnarray}
\lefteqn{\left\{\det \left[\st p+\st AP_L+M\right]
\right\}^*
=\det \left[\st p+\st AP_R+M\right].}\nonumber\\
&&
\end{eqnarray}
That is, the part of the determinant that is odd in $\gamma_5$ is the
phase, and the part that is even in $\gamma_5$ is the magnitude.
Consequently, in a simulation we can effect (\ref{even}) by
replacing the magnitude of the chiral Wilson-Dirac determinant with the
square root of the determinant for Wilson-Dirac particle with
vector-like couplings to the gauge field.

For terms containing an odd number of $\gamma_5$'s, there is clearly no
way to use the preceding trick to eliminate all the $\gamma_5$'s, and,
for such terms, the chiral gauge current is not conserved.  This is not
too surprising since, if we could have eliminated all of the violations
of chiral current conservation, then we would have found a counterterm
that eliminates the Adler-Bardeen-Jackiw anomaly, contrary to the proof
of Adler and Bardeen \cite{adler-bardeen}. Fortunately, it turns out
that {\it all} of the violations of current conservation are proportional
to the anomaly.  Thus, the violations cancel if the fermion species in
the theory satisfy the anomaly-cancellation condition ${\rm Tr}\,
\lambda_a\{\lambda_b,\lambda_c\}=0$, where the $\lambda$'s are the
flavor matrices (or charges) associated with the fermion species.

\subsection{Self-energy and Vertex Corrections}

In general, the graphs associated with the fermion self-energy
correction and fermion--gauge-boson vertex correction lead to six
counterterms: $(Z_1^i-1) \overline\psi\st A P_i\psi$, $\delta
m^i\overline\psi P_i\psi$, and $(Z_2^i-1)\overline\psi ((-i\st \partial)
P_i\psi$, where the index $i$ stands for $L$ or $R$.

In computing the phase of the chiral determinant (that is, the terms
containing an odd number of $\gamma_5$'s), we can choose to drop the
part of the action $S_{WI}$ that corresponds to the gauging of the
Wilson term.  (This does not upset the preceding anomaly-cancellation
argument.)  Then there is a shift symmetry, discussed by Golterman and
Petcher \cite{golterman-petcher}, which guarantees that $(Z_1^R-1)$,
$(Z_2^R-1)$, $\delta m^L$, and $\delta m^R$ vanish. However, it turns
out that $Z_1^L\neq Z_2^L$, so we need a counterterm $(\tilde
Z_1^L-1)S_{NI}$, where $\tilde Z_1^L=Z_1^L/Z_2^L$,
or, equivalently, a counterterm $(\tilde Z_2^L-1)S_{N}$, where $\tilde
Z_2^L=Z_2^L/Z_1^L$.


In a simulation, $\tilde Z_1^L$ (or $\tilde Z_2^L$) must be tuned so
that the renormalized fermion--gauge-boson coupling is the same as the
renormalized triple-gauge-boson coupling.  The dominant contribution to
$\tilde Z_1^L$ comes from the region of large Euclidean loop momenta.
Hence, for asymptotically free theories, $\tilde Z_1^L$ can be computed
in perturbation theory, and it is a finite renormalization.
Unfortunately, $\tilde Z_1^L$ is gauge dependent, so one must gauge fix
in simulations.  However, because $\tilde Z_1^L$ is a local
(perturbative) quantity, it should be insensitive to Gribov ambiguities.
For the terms containing an odd number of $\gamma_5$'s, one can prove a
version of the Adler-Bardeen no-renormalization theorem \cite{no-renorm}
to the effect that, if $\tilde Z_1^L$ is properly adjusted and ${\rm
Tr}\,\lambda_a\{\lambda_b,\lambda_c\}=0$, then the Ward identity for the
complete fermion--gauge-boson vertex is non-anomalous.   That is, the
presence of radiative corrections does not upset the anomaly
cancellation.

In computing the vector-like determinant (whose square root replaces the
magnitude of the chiral determinant), we {\it must} gauge the Wilson
term in order to maintain current conservation.  Hence, there is no
Golterman-Petcher symmetry to protect against a mass counterterm.
However, for a vector-like theory, $Z_1^L=Z_1^R=Z_2^L=Z_2^R$ and $\delta
m^L=\delta m^R=\delta m$.  Since there are no $\tilde Z_1$ or $\tilde
Z_2$ counterterms, we need only tune $\delta m$ (or the hopping
parameter) according to the criterion $m_{renorm}=0$.

\section{SUMMARY}

One can simulate a chiral gauge theory on the lattice through the
following procedure.
\begin{enumerate}

\item Start with an anomaly-free complement of Dirac particles
($\rm{Tr}\, \lambda_a\{\lambda_b,\lambda_c\}=0$) with left-handed
couplings to the gauge field.

\item Fix to a renormalizable gauge.

\item Compute the determinant of each Dirac operator, including in the
Dirac action the naive terms $S_N$ and $S_{NI}$, an ungauged Wilson mass
term $S_{W}$, and a counterterm $(\tilde Z_1^L-1)S_{NI}$ (or $(\tilde
Z_2^L-1)S_N$).

\item Retain the phase of each determinant, but replace its magnitude
with the square root of the determinant of the Dirac operator with a
vector-like coupling to the gauge field.  The vector-like action
includes the naive terms $S_N$ and $S_{NI}$ with $P_L\rightarrow 1$, the
Wilson term $S_W$, and its gauging $S_{WI}$. One must also include a
counterterm for $\delta m$ or a hopping parameter.

\item Tune $\delta m$ so that the physical mass vanishes; tune $\tilde
Z_1^L$ (or $\tilde Z_2^L$) so that the renormalized fermion--gauge-boson
coupling is equal to the renormalized triple-gauge-boson coupling. For
an asymptotically free theory, the critical hopping parameter and
$\tilde Z_1^L$ can be computed in perturbation theory.
\end{enumerate}

%
%

\end{document}